# FINDING A PROMISING VENTURE CAPITAL PROJECT WITH TODIM UNDER PROBABILISTIC HESITANT FUZZY CIRCUMSTANCE

Weike ZHANG[1], Jiang DU[1], Xiaoli TIAN[2,*]

*[1]School of Economics, Sichuan University, Chengdu 610064, China*

*[2] Business School, Sichuan University, Chengdu 610064, China*

**Abstract**. Considering the risk aversion for gains and the risk seeking for losses of venture capitalists, the TODIM has been chosen as the decision-making method. Moreover, group decision is an available way to avoid the limited ability and knowledge etc. of venture capitalists. Simultaneously, venture capitalists may be hesitant among several assessed values with different probabilities to express their real perception because of the uncertain decision-making environment. However, the probabilistic hesitant fuzzy information can solve such problems effectively. Therefore, the TODIM has been extended to probabilistic hesitant fuzzy circumstance for the sake of settling the decision-making problem of venture capitalists in this paper. Moreover, due to the uncertain investment environment, the criteria weights are considered as probabilistic hesitant fuzzy information as well. Then, a case study has been used to verify the feasibility and validity of the proposed TODIM. Also, the TODIM with hesitant fuzzy information has been carried out to analysis the same case. From the comparative analysis, the superiority of the proposed TODIM in this paper has already appeared.

*Corresponding author. E-mail: *tianxiaolitxl@126.com*

**Keywords:** TODIM, probabilistic hesitant fuzzy information, venture capitalist, decision-making, venture capital.

**JEL Classification:** D81, G41.

# Introduction

As Premier Li emphasized, acomprehensively deepen reform is urgent needed to promote transformation and upgrading of Chinese economy in June 7, 2016 during the opening ceremony of World Economic Forum. Because the dazzling growth of economy in China which is driven by investment has become slow in recent years. All of us are eager to find an effective way for the sustainedeconomic growth. Meanwhile, president Xi advocated that we must turn China's economic growth situation from factor-driven and investment-driven to innovation-driven. Innovation is the dynamic of economy, as emphasized by Schumpeter (1934) in innovation theory. Furthermore, innovation promotes the transformation of economy as well and the venture capital (VC) played an important role in such transformation through innovation.

Actually, the reason why VC is so popular those days is that projects with VC-backed get a higher rate of commercialization and are more efficient in innovation than those with non-VC-backed (Dutta&Folta, 2016). While the main characteristic of an entrepreneurial environment is uncertainty, and VC prefer such uncertainty with the intention of gaining extra return through funding a promising project. For instance, the Sequoia Capital, which is the biggest VC firm in the world, has acquired huge profits from early investment of Alibaba, Wanda cinema, Apple computer,etc. However, it also has suffered enormous losses from the investment of Asia-Media digital interactive Co., Ltd(Castilla, 2003).Hence, how to find a promising project is the first and key step for the success of VC and it is also a vital and troubling thing for VCs who operate the VC. Due to the uncertain nature of VC, it is not easy for VCs to select a promising project successfully. Usually,behavioral decision-makingmodel exhibits an important reference in the decision-making process of VCs.

The role of TODIM (TOmada de DecisãoIterativaMulticritério) in behavioral decision has been revealed (Gomes& González, 2012). Researches have already shown that VCs played a primarily role in investment decision (Zacharakis,Mcmullen&Shepherd,2007), and almost all the VCs appeareddifferent risk attitudesfor gains and losses under uncertain environmentthat wererisk aversion for gains and risk seeking for losses(Yazdipour, 2011). Thus, in this paper we adopt TODIMas the reference model for VCs to simulate the risk attitude of them because TODIM is a decision-making model constructed on prospect theory (Kahneman& Tversky, 1979) which is a well-known theory to explain behavioral decision.In particular, the risk aversion for gains and risk seeking for losses are well embodied in TODIM through the dominance function. Additionally, in decision-making process, VCs give assessed values to each project over each criterion under uncertain environment.In addition, group decision is a usual way to overcome the drawbacks of single decision. For example, the limited knowledge, experience, etc. of venture capitalist in single decision may result in an improper decision, whereas, combining the opinions of all VCsare the superiority of group decision in overcoming such drawbacks.Also, the VCs'different experience, diversified educational background, etc.can lead them to conceptualize and understand uncertainty in a different way, and ultimately, give different assessed values.Even if the same venture capitalist may be hesitant among several assessed values and show different degree of hesitation for each assessed value as well,because of uncertain environment and limited ability of him/her. The VCs'different opinions in group decision and each venture capitalist's different degrees of hesitant values reflect that the probability of each possible assessed value may be different.Hence, using the probabilistic hesitant fuzzy set (P-HFS) to portray such evaluation information is reasonable and effective.Takingthe uncertain decision-making environment and risk attitude of VCs into account, it is necessary for us to study the TODIM under probabilistic hesitant fuzzy circumstance and apply this new method to VC field.

The paper intends to provide a decision aid model for VCs to improve their decisions. The main contributions of this paper are: ① The risk attitude of VCs under uncertain VC environmentand fuzzy evaluation information have been considered in this paper simultaneously. ②The combination of TODIM and probabilistic hesitant fuzzy information has been established, and it has been used to solve the decision-making problem of VCs.③Also, the criteria weights are expressed as probabilistic hesitant fuzzy information. ④ The construction of this method will call attention on the combination of fuzzy evaluation information and behavioral decision in VC field.

The rest of this paper is structured as follows. In Section 1, asimple introduction of the concepts and algorithms about probabilistic hesitant fuzzy information and of the steps about classical TODIM will be presented.Then, the decision-making criteria used by VCs are analyzed and explained in Section 2. Also, the method of acquiring criteria weights is established in this section. Next, the new TODIM method under probabilistic hesitant fuzzy circumstance is constructed in Section 3. After that,in Section 4,acase studyisused to exhibit the feasibility of the proposed method. Moreover, the TODIM with hesitant fuzzy information isapplied in comparative analysis todemonstrate the practicability and effectiveness of the proposed methodas well. Finally, some conclusions aredrawn in the last Section.

## 1. Someconcepts and algorithms

*1.1.Probabilistic hesitant fuzzy information*

The probabilistic hesitant fuzzy information is presented byprobabilistic hesitant fuzzy set (P-HFS) which is first proposed in 2014 (Zhu& Xu, 2017).The P-HFS is the extension of hesitant fuzzy set (Torra, 2010) which has been widely used in multi-criteria decision-making (MCDM) (Yu,Zhang & Xu,2013). Also, the dual hesitant fuzzy set has been extended (Zhu,Xu, & Xia,2012; Yu& Li, 2014). However, the P-HFS assigns a probability to every hesitant fuzzy information,which makes a good expression oforiginal perception of decision makers (DMs) for projects.

Let $X$ be a fixed set, a P-HFS on $X$ is expressed as: $H = \{<x, h_x(p_x)> | x \in X\}$. The $h_x(\cdot)$ is a set of some values in [0, 1] and it includes all the possible membership degrees for $x \in X$ to the set $H$. Moreover $p_x$ shows the probability of $h_x(\cdot)$, and $\sum p_x = 1$. For convenience, $h_x(p_x)$ is simply symbolized as $h(p)$ in the following context and it is named as probabilistic hesitant fuzzy element (P-HFE): $h(p) = \{h^t(p^t) | t = 1, 2, \cdots, \#h(p)\}$, where $\#h(p)$ is the number of possible membership degrees and $\sum_{t=1}^{\#h(p)} p^t = 1$. It is interested that $\sum_{t=1}^{\#h(p)} p^t < 1$ is a common situation in real decision-making circumstance, and it is reasonable to assign the incomplete probability information $1 - \sum_{t=1}^{\#h(p)} p^t$ to each $h^t(p^t)$ averagely. Therefore, when $\sum_{t=1}^{\#h(p)} p^t < 1$, the probability of each $h^t(p^t)$ should be reset as $p^t / \sum_{t=1}^{\#h(p)} p^t$, where $\sum_{t=1}^{\#h(p)} (p^t / \sum_{t=1}^{\#h(p)} p^t) = 1$. Then, the comparison of P-HFEs is introduced. As is well-known, score function, variance function and distance measureare the general indexes used for difference analysis between P-HFEs. The score function and variance function are defined as (Ding,Xu, & Zhao,2017):

$$s(h(p)) = \sum_{t=1}^{\#h(p)} p^t h^t(p^t) \quad (1)$$

$$\sigma(h(p)) = \sum_{t=1}^{\#h(p)} p^t (h^t(p^t) - s(h(p)))^2 \quad (2)$$

Based on the Eqs. (1) and (2), the comparative rules are:

1) If $s(h_1(p)) > s(h_2(p))$, then $h_1(p) > h_2(p)$.

2) If $s(h_1(p)) < s(h_2(p))$, then $h_1(p) < h_2(p)$.

3) If $s(h_1(p)) = s(h_2(p))$, then the variance function is used to compare the difference between P-HFEs, that is, $\sigma(h_1(p)) > \sigma(h_2(p))$, then $h_1(p) < h_2(p)$; on the contrary, if $\sigma(h_1(p)) < \sigma(h_2(p))$, then $h_1(p) > h_2(p)$.

The same number of possible membership degrees is the precondition of distance measure between P-HFEs. Let $H_1$ and $H_2$ be two P-HFSs on $X$. If $\#h_1(p)$ is smaller than $\#h_2(p)$, then the number of $\#h_2(p) - \#h_1(p)$ possible membership degrees should be added to $h_1(p)$. Due to the special background of this paper that pursuing huge revenue with high risk is the fundamental characteristic of VC, we should add the biggest membership degree with the probability of zero to $h_1(p)$. It is obvious that such addingrules do not change the values of score function and variance function. Next, we will define the ordered P-HFS. The ordered P-HFS satisfies the following prerequisites:

1) $p^t h^t(p^t) < p^{t+1} h^{t+1}(p^{t+1})$ (ascending order) or $p^t h^t(p^t) > p^{t+1} h^{t+1}(p^{t+1})$ (descending order).

2) If $p^t h^t(p^t) = p^{t+1} h^{t+1}(p^{t+1})$, then the ordering of them is determined by $h^t(p^t)$, the other word, $h^t(p^t) < h^{t+1}(p^{t+1})$ (ascending order) or $h^t(p^t) > h^{t+1}(p^{t+1})$ (descending order) for this situation.

Finally, according to the definition of Hamming distance measure, we define the Hamming distance between the ordered P-HFEs $h_1$ and $h_2$ as:

$$d(h_1, h_2) = \frac{1}{\#h_1(p)} \sum_{t=1}^{\#h_1(p)} | p_1^t h_1^t(p_1^t) - p_2^t h_2^t(p_2^t) |, \quad \#h_1(p) = \#h_2(p). \quad (3)$$

For the sake of simplicity, $H' = \{< x, h'(p') = h''(p'') | t = 1, 2, \cdots, \#h'(p') > | x \in X\}$,

$\sum_{t=1}^{\#h'(p')} h'^t(p') = 1$. Thus, $H'$ and $h'(p')$ represent the normalized P-HFS and P-HFE correspondingly on the basis of the aforementioned rules in the following context.

*1.2. TODIM method*

Prospect theory (Kahneman & Tversky, 1979) which is used to elaborate on the psychological behavior of DMs has been introduced to MCDM method (Gomes & Lima, 1992). Moreover, TODIM(Gomes & Lima, 1991) is an effective and classical MCDM method derived from prospect theory (Kahneman & Tversky, 1979). The fundamental idea of TODIM is to consider the risk attitude of DMs in decision-making process and to measure the relative dominance of each project over the others. Consider the set of venture projects $A = \{A_1, A_2, \cdots, A_n\}$ and attributes $C = \{c_1, c_2, \cdots, c_m\}$. Let $N = \{1, 2, \cdots, n\}$ and $M = \{1, 2, \cdots, m\}$. Then, the steps of classical TODIM method are:

**Step 1.** Obtain the decision information from experts, including the decision matrix $Y = (y_{ij})_{n \times m}$ and criterion weight $\omega$.

$$Y = \begin{pmatrix} y_{11} & \cdots & y_{1m} \\ \vdots & \ddots & \vdots \\ y_{n1} & \cdots & y_{nm} \end{pmatrix} = (y_{ij})_{n \times m}, \omega = (\omega_1, \omega_2, \cdots, \omega_m), \sum_{j=1}^{m} \omega_j = 1.$$

**Step 2.** Transform the decision matrix $Y = (y_{ij})_{n \times m}$ into $Y' = (y'_{ij})_{n \times m}$.

$$y'_{ij} = \begin{cases} y_{ij} & c_j \text{ is benefit criterion} \\ -y_{ij} & c_j \text{ is cost criterion} \end{cases}. \quad (4)$$

**Step 3.** Determine the relative weight $\omega_{jr}$:

$$\omega_{jr} = \frac{\omega_j}{\omega_r}. \quad (5)$$

where $r, j \in M$, $\omega_r = \max(\omega_j \mid j \in M)$ and $c_r$ is called a reference criterion.

**Step 4.** Calculate the dominance of project $A_i$ over $A_k$ ($i, k \in N$):

$$\psi(A_i, A_k) = \sum_{j=1}^{m} \varphi_j(A_i, A_k), \quad (6)$$

where

$$\varphi_j(A_i, A_k) = \begin{cases} \sqrt{\dfrac{\omega_{jr}}{\sum_{j=1}^{m}\omega_{jr}}(y'_{ij} - y'_{kj})} & \text{if } y'_{ij} - y'_{kj} > 0 \\ 0 & \text{if } y'_{ij} - y'_{kj} = 0 \\ \dfrac{-1}{\lambda}\sqrt{\dfrac{\sum_{j=1}^{m}\omega_{jr}}{\omega_{jr}}(y'_{kj} - y'_{ij})} & \text{if } y'_{ij} - y'_{kj} < 0 \end{cases} \quad (7)$$

The parameter $\lambda$ is the attenuation factor of the losses.

**Step 5.** Identify the overall value of project $A_i$:

$$\Omega(A_i) = \frac{\sum_{k=1}^{n}\psi(A_i, A_k) - \min_i\{\sum_{k=1}^{n}\psi(A_i, A_k)\}}{\max_i\{\sum_{k=1}^{n}\psi(A_i, A_k)\} - \min_i\{\sum_{k=1}^{n}\psi(A_i, A_k)\}} \quad i \in N. \quad (8)$$

**Step 6.** Rank the overall value $\Omega(A_i)$, $i \in N$. $A_i$ will be the promising project if $\max(\Omega(A_i), i \in N)$.

The classical TODIM is built on crisp number, and it has been used to solve decision-making problems such as selection of rental residential properties (Gomes& Rangel, 2009), the best option for the destination of the natural gas reserves in Brazil (Gomes,Rangel, &Maranhão,2009), ERP software (Kazancoglu&Burmaoglu, 2013), etc. Also, the criteria interactions have been considered in TODIM (Gomes,Machado, & Rangel,2013). However, the classical TODIM could not successfully express the fuzzinessdecision-making information under uncertainty. Considering the uncertain environment of VC and the risk attitude of VCs, we are dedicated to extendTODIM to probabilistic hesitant fuzzy circumstance for the sake ofhelping VCs to make a better decision. But, before constructing such extension of TODIM, it is significant to discuss the criteria used by VCs in this MCDM problem.

## 2. Discussion about decision-making criteria used by VCs

It is critical for us to know the existing researches about criteria used by VCs to make a better understanding of their decisions and to propose an appropriate method for them as decision aid in this paper.

A majority of literature have indicated that VCs primarily concentrated on the projects' management team, the potential finance of projects, the market conditionsand service or product the project offered when they decided to invest their limited capital (Tyebjee& Bruno, 1984; Macmillan,Zemann, &Subbanarasimha,1987; Hisrich&Jankowicz, 1990; Mason& Stark, 2004; Carpentier&Suret, 2015; Widyanto&Dalimunthe, 2015). In addition, Riquelmeand Rickards (1992) pointed out that managerial experience was a general factor accepted by all VCs to evaluate a project. Furthermore, Franke,Gruber, Harhoff, and Henkel(2008)discovered that project with the member of management team who had experience of the interrelated industry or had a background of crossed education gotthe support of VC more easily. Whereas, acquiring huge revenue is the ultimate goal of VC. Thus, the market with great potential profits and high risk is popular among VC, such as software which is the new technology developed rapidly in recent years and biotechnology which has attracted more and more attention in the past few years. Moreover, from an 11-years period of funded enterprise data in a VC firm, Petty and Gruber (2011)found that VCs considered more about products in final decision as time goes by. However, the management team, the finance situation, the market conditions and the service or product offered by project are the four general criteria accepted by VCs in decision-making process. They are explained as follows:

*(1) Management team( $c_1$ )*

As the CEO of Facebook,Mark Elliot Zuckerberg emphasizes that it is a very important thing to set up a good team for an entrepreneur who wants to start his/her own business.Meanwhile, Paul Graham, the founder of Y Combinator which is a famous business incubator for start-ups in America, also thinks that individualsare the most central part for start-ups.It is clearly recognized by not only investors but also other important stakeholders that a creative and passionate management team will drive the start-ups to the road of success. Furthermore, educational background or experience in the related industry of entrepreneuror management teamand their excellent ability are the decisive factor in the investigation of management team. The VCs prefer the entrepreneurs with higher Emotional Quotient and Intelligence Quotient, with independent thought and an open mind, etc.Hence, it is obvious that the management team of the project shows a significant role for VCs in their investment decision-making process (Widyanto&Dalimunthe, 2015).

*(2)Financial situation ( $c_2$ )*

Lack of capital support leads almost 30% of start-ups to become failing[1]. Although the VCs who provide capital for the project concentrate more on the potential finance of it and acquiring huge profits is the ultimate goal of them, they investigate the current financial situation of the project as well. Also, the pay-back period, return on asset,etc. are considered in investment decision-making process.

*(3) Market condition ( $c_3$ )*

Market demand is dynamic of providing product or service, and it is one of the key factors for the success of start-ups. The reason for the failure of more than forty percent start-ups is lack of effective market demandaccording to the footnote 1. Moreover, market prospect, market growth

---

[1] The results come from the analysis of 101 failing start-ups by CB Insights, a famous data analysis corporate(http://www.gamelook.com.cn/2014/10/185579)

rate, market competition level, etc. are the important aspects in the VCs' decision-making.

*(4) Service or product ( $c_4$ )*

When VCs have chosen a target market, they prefer to investigate whether the product or service provided by the optionalstart-up project is competitive in the aimed market.Also, theacceptability of customers for product or service is an important aspect for VCs in decision-making process because the customers who consume such product or serviceare the basic source of earnings.

In our brief retrospect and explanation, it is known that VCs pay very close attention to management team of project, the financial situation, the market conditionsand the service or product offered by the project. In addition, WidyantoandDalimunthe (2015)has discussed the evaluation criteria in Indonesia via closed-questionnaire and summarized the evaluation criteria around the world. However,the importance degree of each criterionisexpressed as crisp numbers. Whatever, such precise expression seems to be unreasonable owing to the fuzziness of VC environment and the vague perception of VCs.The weight of each criterion is given by individual who is limited in knowledge, experience, etc.Most of time, under the uncertain circumstance, they are hesitantly assign the criterion weight among several values and the hesitant degree of each value is different as well.Moreover, the incomplete criterion weight is common scene in real decision-making situation.Nevertheless, probabilistic hesitant fuzzy information will effectively solve the above problems. Thus, in this paper, the criteria weights are expressed asprobabilistic hesitant fuzzy information.

Suppose $\omega=(\omega_1,\omega_2,\cdots,\omega_m)$ are the weights of the criteria $C=(c_1,c_2,\cdots,c_m)$ respectively and $H = \{<\omega_j, h_{\omega_j}(p_{\omega_j}) = \{h^t_{\omega_j}(p^t_{\omega_j}) | t=1,2,\cdots,\#h(p)\} > | \omega_j \in X\}$, where $\sum_{j=1}^{m} h^{t^+}_{\omega_j}(p^t_{\omega_j}) \leq 1$. The $h^{t^+}_{\omega_j}(p^t_{\omega_j})$

represents the biggest possible weight of $c_j$ in P-HFE $h_{\omega_j}(p_{\omega_j})$. Based on the definition of score function of P-HFE, $s(h_{\omega_j}(p_{\omega_j}))$ represents the expected criterion weight. Therefore, it is appropriate to treat the weight of $c_j$ as:

$$\omega_j = s(h_{\omega_j}(p_{\omega_j})) = \sum_{t=1}^{\#h_{\omega_j}(p_{\omega_j})} p_{\omega_j}^t h_{\omega_j}^t(p_{\omega_j}^t). \qquad (9)$$

It is apparent that $\sum_{j=1}^{m}\omega_j \neq 1$ is a general phenomenon. Then, we should normalize the weight of criterion as:

$$\omega'_j = \frac{\omega_j}{\sum_{j=1}^{m}\omega_j}, \qquad (10)$$

where $\sum_{j=1}^{m}\omega'_j = 1$.

Since the weights of criteria are obtained, the new TODIM with probabilistic hesitant fuzzy information will be established in the next section.

## 3. Construction a new TODIM under probabilistic hesitant fuzzy circumstance

The TODIM, an effective MCDM method to simulate the risk attitude of DMs, has been applied in numerous fields and extended to diverse fuzzy circumstance such as intuitionistic fuzzy (Krohling, Pacheco, &Siviero, 2013), interval intuitionistic fuzzy (Krohling& Pacheco, 2014), hesitant fuzzy (Zhang& Xu, 2014), hesitant fuzzy linguistic (Wei, Ren, &Rodr ǵuez, 2015). Moreover, in regard to the same problem, sometimes the form of decision-making information may be various. Hence, the TODIM has been extended to the circumstance of hybrid information (Fan, Zhang, Chen, & Liu, 2013). However, none of the extension of TODIM is concerned with probabilistic hesitant fuzzy information. But probabilistic hesitant fuzzy information can successfully solve the group decision-making problem and perfectlydeal with the situation that venture capitalistis hesitant among several evaluation values with different probabilities. Therefore, in this paper we are dedicated to construct a new TODIM with probabilistic hesitant fuzzy information that seems to be missing discussed in the previous literature and apply the proposed method to select a promising project for VCs.

The steps of the new TODIM under probabilistic hesitant fuzzy circumstanceare:

**Step 1.** Identify the decision-making problem, and then, confirm the optional projects $A=\{A_1, A_2, \cdots, A_n\}$ and the decision-making criteria $C=\{c_1, c_2, \cdots, c_m\}$.

**Step 2.** Acquire the original evaluation information of projectsfrom VCs as:

$$Y = \begin{pmatrix} h_{11}(p_{11}) & \cdots & h_{1m}(p_{1m}) \\ \vdots & \ddots & \vdots \\ h_{n1}(p_{n1}) & \cdots & h_{nm}(p_{nm}) \end{pmatrix} = (h_{ij}(p_{ij}))_{n \times m}, W = (h_{\omega_1}(p_{\omega_1}), h_{\omega_2}(p_{\omega_2}), \cdots, h_{\omega_m}(p_{\omega_m})), \quad (11)$$

where $i \in N$, $j \in M$; $h_{ij}(p_{ij})$ is the P-HFE and represents the assessed value for project $A_i$ under criterion $c_j$; $h_{\omega_j}(p_{\omega_j})$ shows weightinformation of criterion $c_j$.

**Step 3.** Standardize evaluation information, including normalizing the evaluation matrix according to Section 1.1. and calculating criteria weights on the basis ofEqs. (9) and (10):

$$Y' = \begin{pmatrix} h'_{11}(p'_{11}) & \cdots & h'_{1m}(p'_{1m}) \\ \vdots & \ddots & \vdots \\ h'_{n1}(p'_{n1}) & \cdots & h'_{nm}(p'_{nm}) \end{pmatrix} = (h'_{ij}(p'_{ij}))_{n \times m}, \omega' = (\omega'_1, \omega'_2, \cdots, \omega'_m), \quad (12)$$

where $\sum_{j=1}^{m} p'_{ij} = 1$ ($i \in N$) and $\sum_{j=1}^{m} \omega'_j = 1$.

**Step 4.** Calculate the relative criterion weight $\omega'_{jr}$ according to reference criterion:

$$\omega'_{jr} = \frac{\omega'_j}{\omega'_r}, \qquad (13)$$

where $\omega'_r$ is the weight of reference criterion $c_r$, and $\omega'_r = \max(\omega'_j | j \in M)$.

**Step 5.** Determine the relative dominance of gains or losses for project $A_i$ to $A_k$ under criterion $c_j$. It is represented as $\vartheta_j(A_i, A_k)$ which includes benefit criteria and cost criteria. Hence, if $c_j$ is a benefit criterion, the relative dominance will be:

$$\vartheta_j^B(A_i, A_k) = \begin{cases} \sqrt{\frac{\omega'_{jr}}{\sum_{j=1}^{m} \omega'_{jr}}} d(h'_{ij}(p'_{ij}), h'_{kj}(p'_{kj})) & h'_{ij}(p'_{ij}) > h'_{kj}(p'_{kj}) \\ 0 & h'_{ij}(p'_{ij}) = h'_{kj}(p'_{kj}) \\ -\frac{1}{\lambda} \sqrt{\frac{\sum_{j=1}^{m} \omega'_{jr}}{\omega'_{jr}}} d(h'_{ij}(p'_{ij}), h'_{kj}(p'_{kj})) & h'_{ij}(p'_{ij}) < h'_{kj}(p'_{kj}) \end{cases}, (14)$$

if $c_j$ is a cost criterion, the relative dominance will be:

$$\vartheta_j^C(A_i, A_k) = \begin{cases} -\frac{1}{\lambda} \sqrt{\frac{\sum_{j=1}^{m} \omega'_{jr}}{\omega'_{jr}}} d(h'_{ij}(p'_{ij}), h'_{kj}(p'_{kj})) & h'_{ij}(p'_{ij}) > h'_{kj}(p'_{kj}) \\ 0 & h'_{ij}(p'_{ij}) = h'_{kj}(p'_{kj}) \\ \sqrt{\frac{\omega'_{jr}}{\sum_{j=1}^{m} \omega'_{jr}}} d(h'_{ij}(p'_{ij}), h'_{kj}(p'_{kj})) & h'_{ij}(p'_{ij}) < h'_{kj}(p'_{kj}) \end{cases}. (15)$$

**Step 6.** Aggregate the dominance of project $A_i$ to $A_k$:

$$\theta(A_i, A_k) = \sum_{j=1}^{m} \vartheta_j(A_i, A_k). \quad (16)$$

**Step 7.** Collect the overall dominance of project $A_i$:

$$O(A_i) = \frac{\sum_{k=1}^{n} \theta(A_i, A_k) - \min_i \{\sum_{k=1}^{n} \theta(A_i, A_k)\}}{\max_i \{\sum_{k=1}^{n} \theta(A_i, A_k)\} - \min_i \{\sum_{k=1}^{n} \theta(A_i, A_k)\}} \quad \forall i, k \in N. \quad (17)$$

**Step 8.** Rank the $O(A_i)$. The best project will be the one which has the biggest $O(A_i)$.

The visual procedure of the proposed TODIM isshown in Figure 1:

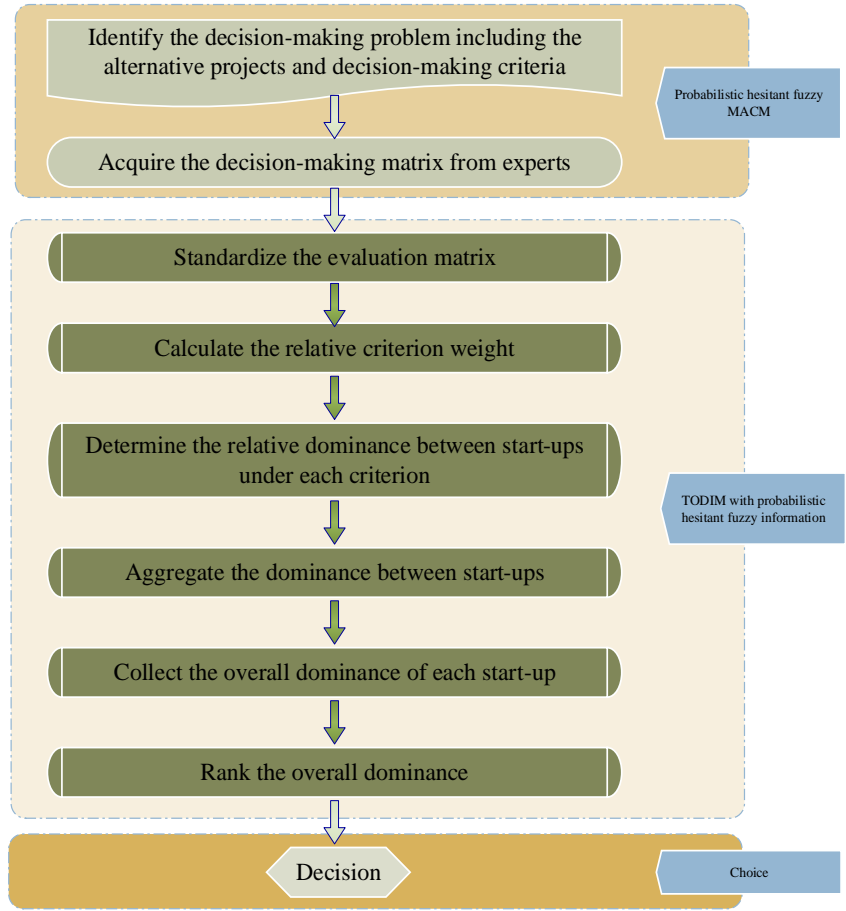

Figure 1. The decision-making process of TODIM with probabilistic hesitant fuzzy information

Up to now, the decision-making method of TODIM under probabilistic hesitant fuzzy circumstance has been constructed. Then, the feasibility and usefulness of this method will be exhibited in Section 4 through a case study. A comparative analysis between the proposed TODIM and TODIM with hesitant fuzzy information will be presented as well.

## 4. A case study

Talent person is a driving force for national development, whereas, education is the most fundamental way to obtain the talent person. Moreover, in China, it is a widespread phenomenon that all the parents hope their children have a bright future. Therefore, with the development of economy and the improvement of living standards, more and more parents pay much attention to the education problem of their children. In particular, the completely releasedpolicy of a two-child per family will drastically increase the number of children in China. It is obvious that the educational industry will welcome a bright prospect.

Now, there are numerous of educational institutions in China, and they focus on early education,instruction after class, talent cultivation such as host, dance, musical instruments, etc. Also, as the popularization of internet technology, various online courses with less charge or free have been prevalent, such as the classroom of Tencent[2], the public course of Netease[3], etc.Impacted by the internet, offline educational institution has been actively involved in combining the experience and superiority of offline with online technology to be more competitive.Considering the traditional education ideas of Chinese parents that academic achievement comes first and the drastic academic competition among student, in this paper, we concentrate on the optionaleducational institutions which specialize in the instruction after class in the stage of elementary school, middle school and high school. Therefore, after careful screening, four famous educational institutions have been left to be further investigation.

Shanghai Yimi Education Techology Co., Ltd.[4] is aninternet educational institutionfocused on the instruction after class fromelementary school to high school. For the sake of developing distance education, it combines the excellent offline educational resources withleading internet technology. Moreover, it has been the navigator of self-operated mobile online education and has honored the brand enterprise of Chinese internet education in May 2016.Beijing Zhenguanyu Technology Co., Ltd. provides many electronic learning products named Yuantiku (an APP with intelligent question bank covering all the courses in middle school and high school for students to practice), Yuanfudao

---
[2] https://ke.qq.com/
[3] http://open.163.com/cuvocw/
[4] http://www.1mifudao.com/

(anAPPmade all the students receive mentoring from famous teachers nationwide through live streaming), Banmasusuan(an APP designed for children between 5 to 10 years old in order to quickly improve their computing ability through everyday fun breakthrough and PK practice), etc. It devotes to the online distance education and has achieved good education effects. Qinxue (Beijing) Network Education Technology Co. Ltd.[5] is a training institution including the Qinxueyun (an intelligent education platform covering elementary school to high school), the high-quality and personalized after-school tutoring center covering one toone service, interactive small-sized class, etc.In addition, it provides the professional training for the students who want to participant in independent recruitment, including draw up study plan, selecting school, application, preparation of materials and interview, etc. Puxin Education Technology Group Co. Ltd.[6]has been founded in 2014,specialized in after school tutorial programs from elementary school to high school. It is leaded by an elite team of senior executives with an average 15 years of education management experience. The service of Puxin coversall subjects such as English, Chinese, mathematics, academic testing and assessments, academic tutoring for art students, independent student recruitment, boarding school program, all-day program, after school program, etc. Those four educational institutions are represented as $A_1$, $A_2$, $A_3$, $A_4$ respectively.Then, a group of expertized investors who are familiar with the educational industry have been invited to investigatethe special projects of those four educational institutions. After furious discussion about all the aspects of those projects, a consistentsuggestion has been reached. Hereby, the decision-making process with TODIM under probabilistic hesitant fuzzy circumstance and under hesitant fuzzy circumstance will be exhibited in Section 4.1and Section 4.2correspondingly:

---

[5] http://www.qinxue100.com/
[6] http://www.pxjy.com/

## 4.1. TODIM with probabilistic hesitant fuzzy information

According to the steps established in Section 4, the screening processesare:

**Step 1**. As is mentioned above, the decision-making problem is to find out a promising project of the four educational institutions based on the criteria $C = (c_1, c_2, c_3, c_4)$ discussed in Section 2 and the investors agree with those criteria.

A group of professional investors are asked to give their evaluation information from 0 to 1. If an investor thinks the membership degree of $A_1$ to the optimal management team is 0.55 with the probability of 0.12, then the evaluation information will be $\{0.55(0.12)\}$. For the purpose of distinguishing with the probability of each assessed value, we have multiplied the assessed value of project $A_i$ over criterion $c_j$ by 100 in the following context while the weight information is not changed. Hence, the evaluation information is changed as $\{55(0.12)\}$. Additionally, the decision-making criteria such as management team ($c_1$), financial situation ($c_2$), marketing condition ($c_3$), service or product ($c_4$) are all benefit ones.

**Step 2**. The evaluation information given by the invited investors are obtained as:

$$Y = \begin{array}{c} \\ A_1 \\ A_2 \\ A_3 \\ A_4 \end{array} \begin{pmatrix} c_1 & c_2 & c_3 & c_4 \\ \{55(0.22),68(0.51),73(0.27)\} & \{60(0.61),66(0.39)\} & \{62(0.69),68(0.21)\} & \{64(0.66),72(0.32)\} \\ \{62(0.28),77(0.63)\} & \{68(0.29),77(0.71)\} & \{60(0.18),73(0.21),85(0.61)\} & \{77(0.60),88(0.36)\} \\ \{63(0.32),71(0.48),77(0.12)\} & \{66(0.48),71(0.52)\} & \{68(0.59),74(0.32)\} & \{71(0.53),78(0.22),81(0.25)\} \\ \{67(0.49),72(0.44)\} & \{62(0.55),69(0.45)\} & \{67(0.61),71(0.26)\} & \{68(0.36),73(0.41),79(0.15)\} \end{pmatrix}$$

$W = (\{0.34(0.68), 0.40(0.32)\}, \{0.09(0.39), 0.11(0.61)\}, \{0.19(0.56), 0.22(0.44)\}, \{0.21(0.43), 0.27(0.57)\})$.

**Step 3**. The evaluation information is normalized as:

$$Y = \begin{array}{c} \\ A_1 \\ A_2 \\ A_3 \\ A_4 \end{array} \begin{pmatrix} c_1 & c_2 & c_3 & c_4 \\ \{55(0.22),68(0.51),73(0.27)\} & \{60(0.61),66(0.39)\} & \{62(0.77),68(0.23),68(0)\} & \{64(0.67),72(0.33),72(0)\} \\ \{62(0.31),77(0.69),77(0)\} & \{68(0.29),77(0.71)\} & \{60(0.18),73(0.21),85(0.61)\} & \{77(0.625),88(0.375),88(0)\} \\ \{63(0.35),71(0.52),77(0.13)\} & \{66(0.48),71(0.52)\} & \{68(0.65),74(0.35),74(0)\} & \{71(0.53),78(0.22),81(0.25)\} \\ \{67(0.53),72(0.47),72(0)\} & \{62(0.55),69(0.45)\} & \{67(0.70),71(0.30),71(0)\} & \{68(0.39),73(0.45),79(0.16)\} \end{pmatrix}$$

$W = (0.395, 0.112, 0.224, 0.269)$.

**Step 4**. The relative criteria weights are calculated as (Table 1):

**Table 1. Relative criteria weights**

| $\omega_{1r}$ | $\omega_{2r}$ | $\omega_{3r}$ | $\omega_{4r}$ |
|---|---|---|---|
| 1 | 0.28 | 0.57 | 0.68 |

**Step 5**. The relative dominance between projects over each criterion are determined by Eqs.(14) and (15) (Table 2):

**Table 2. Relative dominance between projects for each criterion**

| Criterion / Dominance | $c_1$ | $c_2$ | $c_3$ | $c_4$ | Criterion / Dominance | $c_1$ | $c_2$ | $c_3$ | $c_4$ |
|---|---|---|---|---|---|---|---|---|---|
| $\vartheta_j(A_1, A_2)$ | −2.29 | −4.60 | −2.15 | −1.89 | $\vartheta_j(A_3, A_1)$ | 0.93 | 0.59 | 1.00 | 2.20 |
| $\vartheta_j(A_1, A_3)$ | −1.05 | −2.34 | −2.00 | −3.64 | $\vartheta_j(A_3, A_2)$ | −2.20 | −5.10 | −2.94 | −3.49 |
| $\vartheta_j(A_1, A_4)$ | −2.12 | −2.61 | −1.32 | −2.55 | $\vartheta_j(A_3, A_4)$ | −2.00 | 0.44 | 0.76 | 1.57 |
| $\vartheta_j(A_2, A_1)$ | 2.03 | 1.16 | 1.08 | 1.14 | $\vartheta_j(A_4, A_1)$ | 1.89 | 0.66 | 0.66 | 1.54 |
| $\vartheta_j(A_2, A_3)$ | 1.96 | 1.29 | 1.48 | 2.11 | $\vartheta_j(A_4, A_2)$ | −2.34 | −5.29 | −2.52 | −2.92 |
| $\vartheta_j(A_2, A_4)$ | 2.08 | 1.34 | 1.27 | 1.77 | $\vartheta_j(A_4, A_3)$ | 1.78 | −1.74 | −1.51 | −2.60 |

Note: The $\vartheta_j(A_i, A_i) = 0$ is not exhibited in the table, and so does $\theta_j(A_i, A_i) = 0$ in the next context.

**Step 6**. The dominance between projects are aggregated as (Table 3):

**Table 3. Dominance between projects**

| | | | | | | | |
|---|---|---|---|---|---|---|---|
| $\theta_j(A_1, A_2)$ | −10.92 | $\theta_j(A_2, A_1)$ | 5.42 | $\theta_j(A_3, A_1)$ | 4.74 | $\theta_j(A_4, A_1)$ | 4.76 |
| $\theta_j(A_1, A_3)$ | −9.04 | $\theta_j(A_2, A_3)$ | 6.84 | $\theta_j(A_3, A_2)$ | −13.73 | $\theta_j(A_4, A_2)$ | −13.08 |
| $\theta_j(A_1, A_4)$ | −8.61 | $\theta_j(A_2, A_4)$ | 6.46 | $\theta_j(A_3, A_4)$ | 0.77 | $\theta_j(A_4, A_3)$ | −4.07 |

**Step 7**. The overall dominance of each project is collected as (Table 4):

**Table 4. Overall dominance of projects**

| $O(A_1)$ | $O(A_2)$ | $O(A_3)$ | $O(A_4)$ |
|---|---|---|---|
| 0 | 1 | 0.43 | 0.34 |

**Step 8**. According to Table 4, the ranking result is: $O(A_2) > O(A_3) > O(A_4) > O(A_1)$. Thus, $A_2 \succ A_3 \succ A_4 \succ A_1$.

*4.2. TODIM with hesitant fuzzy information*

Hesitant fuzzy information describes the hesitant situation of DMs without probability. Many researches have focus on TODIM with hesitant fuzzy information (Zhang& Xu, 2014; Zhang& Xu, 2017). Also, based on Choquet integral, the hesitant fuzzy TODIM has been studied (Tan,Jiang, & Chen,2015; Peng,Wang, Zhou, & Chen,2015). Becauseprobabilistic hesitant fuzzy information is the enhanced version of hesitant fuzzy informationin describing the real decision-making situation, it is reasonable for us to adopt TODIM under hesitant fuzzy circumstance to make a comparative analysis with proposed TODIM in this paper.The detailed steps of TODIM under hesitant fuzzy circumstanceare shown as below:

**Step 1**. Understand the decision-making problem.

**Step 2**. Acquire the evaluation information under hesitant fuzzy circumstance. In order to bemuch more comparable and visualized, the TODIM with hesitant fuzzy information has been used to analysis the aforementioned case. Therefore, the evaluation informationis the same as step 2 in Section 4.1, whereas, the difference is that the hesitant fuzzy information without probability.

$$Y = \begin{matrix} & C_1 & C_2 & C_3 & C_4 \\ x_1 & \{55,68,73\} & \{60,66\} & \{62,68\} & \{64,72\} \\ x_2 & \{62,77\} & \{68,77\} & \{60,73,85\} & \{77,88\} \\ x_3 & \{63,71,77\} & \{66,71\} & \{68,74\} & \{71,78,81\} \\ x_4 & \{67,72\} & \{62,69\} & \{67,71\} & \{68,73,79\} \end{matrix}$$

$$W = (\{0.34, 0.40\}, \{0.09, 0.11\}, \{0.19, 0.22\}, \{0.21, 0.27\}).$$

**Step 3.** Calculate the relative criterion weight $\omega_{jr}$:

$$\omega = \frac{\omega_j}{\omega_r} = \frac{s(h_{\omega_j})}{s(h_{\omega_r})}, \quad (18)$$

where $j, r \in M$, $s(h_{\omega_i})$ is the score function of hesitant fuzzy element $h_{\omega_j}$ (Xia, Xu, & Chen, 2013):

$$s(h_{\omega_i}) = \frac{1}{\#h_{\omega_i}} \sum_{t=1}^{\#h_{\omega_i}} h_{\omega_i}^t. \quad (19)$$

Hence, the relative criteria weights will be (Table 5):

**Table 5. Relative criteria weights**

| $\omega_{1r}$ | $\omega_{2r}$ | $\omega_{3r}$ | $\omega_{4r}$ |
|---|---|---|---|
| 1 | 0.27 | 0.55 | 0.65 |

**Step 4.** Obtain the relative dominance of project $A_i$ over $A_k$ under criterion $c_j$. It is represented as $\vartheta_j(A_i, A_k)$ which includes the benefit criteria and cost criteria. Thus, if $c_j$ is benefit criterion, the relative dominance will be:

$$\vartheta_j^B(A_i, A_k) = \begin{cases} \sqrt{\dfrac{\omega_{jr}}{\sum_{j=1}^m \omega_{jr}} d(h_{ij}, h_{kj})} & h_{ij} > h_{kj} \\ 0 & h_{ij} = h_{kj} \\ -\dfrac{1}{\lambda}\sqrt{\dfrac{\sum_{j=1}^m \omega_{jr}}{\omega_{jr}} d(h_{ij}, h_{kj})} & h_{ij} < h_{kj} \end{cases}, \quad (20)$$

if $c_j$ is cost criterion, the relative dominance will be:

$$\vartheta_j^C(A_i, A_k) = \begin{cases} -\dfrac{1}{\lambda}\sqrt{\dfrac{\sum_{j=1}^m \omega_{jr}}{\omega_{jr}} d(h_{ij}, h_{kj})} & h_{ij} > h_{kj} \\ 0 & h_{ij} = h_{kj} \\ \sqrt{\dfrac{\omega_{jr}}{\sum_{j=1}^m \omega_{jr}} d(h_{ij}, h_{kj})} & h_{ij} < h_{kj} \end{cases}, \quad (21)$$

where the Hamming distance measure of hesitant fuzzy element is(Xu& Xia, 2011):

$$d(h_{ij}, h_{kj}) = \frac{1}{\#h_{ij}} \sum_{t=1}^{\#h_{ij}} |h_{ij}^t - h_{kj}^t|, \quad \#h_{ij} = \#h_{kj}. \quad (22)$$

As the comparative rules mentioned in Section 1.1, the difference of hesitant fuzzy element comes from score function (19) and various function (23) (Liao,Xu, & Xia,2014):

$$\sigma(h_{ij}) = \frac{1}{\#h_{ij}} \sqrt{\sum_{\forall h_{ij}^t \in h_{ij}} (h_{ij}^t - s(h_{ij}))^2}. \quad (23)$$

Therefore, the relative dominance $\vartheta_j(A_i, A_k)$ will be (Table 6):

**Table 6. Relative dominance under each criterion**

| Criteria Dominance | $c_1$ | $c_2$ | $c_3$ | $c_4$ | Criteria Dominance | $c_1$ | $c_2$ | $c_3$ | $c_4$ |
|---|---|---|---|---|---|---|---|---|---|
| $\vartheta_j(A_1, A_2)$ | −1.80 | −4.14 | −2.66 | −3.36 | $\vartheta_j(A_3, A_1)$ | 1.42 | 0.78 | 1.16 | 1.39 |
| $\vartheta_j(A_1, A_3)$ | −1.56 | −3.15 | −2.30 | −2.35 | $\vartheta_j(A_3, A_2)$ | 0.97 | −2.69 | −2.42 | −2.40 |
| $\vartheta_j(A_1, A_4)$ | −1.66 | −2.13 | −1.80 | −1.74 | $\vartheta_j(A_3, A_4)$ | 1.16 | 0.57 | 0.72 | 0.94 |
| $\vartheta_j(A_2, A_1)$ | 1.64 | 1.02 | 1.34 | 1.98 | $\vartheta_j(A_4, A_1)$ | 1.51 | 0.52 | 0.91 | 1.02 |
| $\vartheta_j(A_2, A_3)$ | −1.07 | 0.66 | 1.22 | 1.42 | $\vartheta_j(A_4, A_2)$ | 1.42 | −3.56 | −2.60 | −2.88 |
| $\vartheta_j(A_2, A_4)$ | −1.56 | 0.87 | 1.31 | 1.70 | $\vartheta_j(A_4, A_3)$ | −1.28 | −2.33 | −1.43 | −1.58 |

**Step 5.** Work out the dominance according to Eq. (16) (Table 7):

**Table 7. Dominance between projects**

| | | | | | | | |
|---|---|---|---|---|---|---|---|
| $\theta_j(A_1, A_2)$ | −11.97 | $\theta_j(A_2, A_1)$ | 5.98 | $\theta_j(A_3, A_1)$ | 4.74 | $\theta_j(A_4, A_1)$ | 3.97 |
| $\theta_j(A_1, A_3)$ | −9.37 | $\theta_j(A_2, A_3)$ | 2.23 | $\theta_j(A_3, A_2)$ | −6.54 | $\theta_j(A_4, A_2)$ | −7.61 |
| $\theta_j(A_1, A_4)$ | −7.32 | $\theta_j(A_2, A_4)$ | 2.32 | $\theta_j(A_3, A_4)$ | 3.39 | $\theta_j(A_4, A_3)$ | −6.62 |

**Step 6.** Collect the overall dominance of project $A_i$ on the basis of Eq. (17) (Table 8):

**Table 8. Overall dominance of each project**

| $O(A_1)$ | $O(A_2)$ | $O(A_3)$ | $O(A_4)$ |
|---|---|---|---|
| 0 | 1 | 0.77 | 0.47 |

**Step 7.** Rank $O(A_i)$, then, $O(A_2) > O(A_3) > O(A_4) > O(A_1)$. Hence, $A_2 \succ A_3 \succ A_4 \succ A_1$.

*4.3. The comparison of the two methods*

InSection 4.2 and Section 4.3, the results of TODIM under probabilistic hesitant fuzzy circumstance and under hesitant fuzzy circumstance has been worked out correspondingly as showing in Table 9.

**Table 9. Ranking results of TODIM with different information**

| Ranking results / TODIM with different information | $O(A_1)$ | $O(A_2)$ | $O(A_3)$ | $O(A_4)$ |
|---|---|---|---|---|
| Probabilistic hesitant fuzzy information | 4 | 1 | 2 | 3 |
| Hesitant fuzzy information | 4 | 1 | 2 | 3 |

It is easy enough to recognize that the ranking results of the two are the same. Even so, the proposed TODIM with probabilistic hesitant fuzzy information depicts the differentprobability of each possible assessed value while TODIM with hesitant fuzzy information considers that the probability of each possible assessed value is the same. In particularly, the proposed TODIM can reflect the different opinions of all the VCs in group decision-making situation. For example, ten VCs have been invited to evaluate the market potential of a start-up project. Two VCs give 72, three of them assign 78, four VCs think 82 is reasonable and only one venture capitalist grades 90. In this situation, the real evaluation information will be translated as probabilistic hesitant fuzzy information $\{71(0.2), 78(0.3), 82(0.4), 90(0.1)\}$. But if we use hesitant fuzzy information to express this situation, it will be $\{71, 78, 82, 90\}$ and the number of VCs has been ignored in hesitant fuzzy

information.Furthermore, the proposed TODIM also includes each venture capitalist' different degree of hesitation. For instance, if the first venture capitalist thinks that the market potential will be 71 with the probability of 65% or be 78 with the probability of 35%, then the comprehensive expression of such evaluation information will be $\{71(0.65), 78(0.35)\}$ as probabilistic hesitant fuzzy information or be $\{71, 78\}$ as hesitant fuzzy information.From here we see that hesitant fuzzy information can not express the above phenomenon in real decision-making situation, and itmay create information distortion and finally lead to improper decision. Thus, the proposed TODIM is superior than TODIM under hesitant fuzzy circumstance. Also, it includes more original information than the others extension of TODIM. Furthermore, thedecision-making with proposed model in this paper is helpfulforVCs.

## Conclusions

This paper adopts TODIM, which is a useful technology developed from prospect theory, to portray the VCs' risk attitude that is risk aversion for gains and risk seeking for losses. Furthermore, considering the uncertain circumstance of VC and the vague perception of VCs, it has been extended to probabilistic hesitant fuzzy circumstance and the criteria weights have also been considered as probabilistic hesitant fuzzy information. The detailed steps of the extended TODIM have been given. Also, the decision-making problem has been presented in order to exhibit the reasonability and superiority through the comparison of the proposed TODIM and the TODIM with hesitant fuzzy information. Although there is no difference in ranking results, the proposed TODIM included more original decision-making information is superior than the others. Moreover,it is particularly appropriate the group decision-making of VCs and the different hesitant degree among several assessed values of VCs.

The application of the proposed method in this paper demonstrates the need to be able to model fuzzy information related to VC. This is indeed accomplished in this paper by making use of TODIM under probabilistic hesitant fuzzy circumstance. Moreover, a generalized TODIM proposed by Llamazares (2018) is interesting and deserved to extended under fuzzy circumstance. Furthermore, the proposed method will promote the combination of behavioral decision and fuzzy information. Therefore, as our expectation, more and more researches will be done about prospect theory, regret theory, overconfidence theory, etc. with fuzzy information in the near future.

## Acknowledgments

The research was supported by the National Social Science Foundation of China (Nos. 14BJY176) and the Fundamental Research Funds for the CentralUniversities.

## References

Carpentier, C., &Suret, J.M. (2015). Angel group members'decision process and rejection criteria: a longitudinal analysis. *Journal of Business Venturing, 30*(6), 808-821.https://doi.org/10.1016/j.jbusvent.2015.04.002

Castilla, E.J. (2003). Networks of venture capital firms in Silicon Valley. *International Journal of Technology Management, 25*(25), 113-135.https://doi.org/10.1504/IJTM.2003.003093

Ding, J., Xu, Z.S., &Zhao, N. (2017). An interactive approach to probabilistic hesitant fuzzy multi-attribute group decision making with incomplete weight information. *Journal of Intelligent and Fuzzy Systems, 32*(3), 2523-2536.https://doi.org/10.3233/JIFS-16503

Dutta, S., &Folta, T.B. (2016). A comparison of the effect of angels and venture capitalists on innovation and value creation. *Journal of Business Venturing, 31*(1), 39-54.https://doi.org/10.1016/j.jbusvent.2015.08.003

Fan, Z.P., Zhang, X., Chen, F.D., &Liu, Y. (2013). Extended TODIM method for hybrid multiple attribute decision making problems. *Knowledge-Based Systems, 42*(2),

40-48.https://doi.org/10.1016/j.knosys.2012.12.014

Franke, N., Gruber, M., Harhoff, D., &Henkel, J. (2008). Venturecapitalists'evaluations of start-up teams: trade-offs, knock-out criteria, and the impact of VC experience. *Entrepreneurship Theory and Practice, 32*(3), 459-483.https://doi.org/10.1111/j.1540-6520.2008.00236.x

Gomes, L.F.A.M., &González, X.I. (2012). Behavioral multi-criteria decision analysis: further elaborations on the TODIM method. *Foundations of Computing and Decision Sciences, 37*(1), 3-8.https://doi.org/10.2478/v10209-011-0001-1

Gomes, L.F.A.M., &Lima, M.M.P.P. (1991). TODIM: basic and application to multicriteria ranking of projects with environmental impacts. *Foundations of Computing and decision Sciences, 16*(3-4), 113-127.

Gomes, L.F.A.M., &Lima, M.M.P.P. (1992).From modelling individual preferences to multicriteria ranking of discrete alternatives: a look at Prospect Theory and the additive difference model. *Foundations of Computing and Decision Sciences, 17*(3), 171-184.

Gomes, L.F.A.M., Machado, M.A.S., &Rangel, L.A.D. (2013). Behavioral multi-criteria decision analysis: The TODIM method with criteria interactions. *Annals of Operations Research, 211*(1),531-548.https://doi.org/10.1007/s10479-013-1454-9

Gomes, L.F.A.M., &Rangel, L.A.D. (2009). An application of the TODIM method to the multicriteria rental evaluation of residential properties. *European Journal of Operational Research, 193*(1), 204-211.https://doi.org/10.1016/j.ejor.2007.10.046

Gomes, L.F.A.M., Rangel, L.A.D., &Maranhão, F.J.C. (2009).Multicriteria analysis of natural gas destination in Brazil: an application of the TODIM method. *Mathematical and Computer Modelling, 50*(1), 92-100.https://doi.org/10.1016/j.mcm.2009.02.013

Hisrich, R.D., &Jankowicz, A.D. (1990). Intuition in venture capital decisions: an exploratory study using a new technique. *Journal of Business Venturing, 5*(1), 49-62.https://doi.org/10.1016/0883-9026(90)90026-P

Kahneman, D., &Tversky, A. (1979). Prospect theory: analysis of decision under risk. *Econometria, 47*, 263-291.https://doi.org/10.1142/9789814417358_0006

Kazancoglu, Y., &Burmaoglu, S. (2013). ERP software selection with MCDM: application of TODIM method. *International Journal of Business Information Systems, 13*(4),

435-452.https://doi.org/10.1504/IJBIS.2013.055300

Krohling, R.A., &Pacheco, A.G.C. (2014). Interval-valued intuitionistic fuzzy TODIM. *Procedia Computer Science, 31*, 236-244.https://doi.org/10.1016/j.procs.2014.05.265

Krohling, R.A., Pacheco, A.G.C., &Siviero, A.L.T. (2013). IF-TODIM: an intuitionistic fuzzy TODIM to multi-criteria decision making. *Knowledge-Based Systems, 53*(9), 142-146.https://doi.org/10.1016/j.knosys.2013.08.028

Liao, H.C., Xu, Z.S., &Xia, M.M. (2014). Multiplicative consistency of hesitant fuzzy preference relation and its application in group decision making. *International Journal of Information Technology and Decision Making, 13*(01), 47-76.https://doi.org/10.1142/S0219622014500035

Llamazares, B. (2018). An analysis of the generalized TODIM method. *European Journal of Operational Research, 269*, 1041-1049.https://doi.org/10.1016/j.ejor.2018.02.054

Macmillan, I.C., Zemann, L., &Subbanarasimha, P.N. (1987). Criteria distinguishing successful from unsuccessful ventures in the venture screening process. *Journal of Business Venturing, 2*(2), 123-137.https://doi.org/10.1016/0883-9026(87)90003-6

Mason, C., &Stark, M. (2004). What do investors look for in a business plan? A comparison of the investment criteria of bankers, venture capitalists, and business angles. *International Small Business Journal, 22*(3), 227-248.https://doi.org/10.1177/0266242604042377

Peng, J.J., Wang, J.Q., Zhou, H., &Chen, X.H. (2015). A multi-criteria decision-making approach based on TODIM and Choquet integral within a multiset hesitant fuzzy environment. *Applied Mathematics and Information Sciences, 9*(4), 2087-2097.http://dx.doi.org/10.12785/amis/090448

Petty, J.S., &Gruber, M. (2011). "In pursuit of the real deal": a longitudinal study of VC decision making. *Journal of Business Venturing, 26*(2), 172-188.https://doi.org/10.1016/j.jbusvent.2009.07.002

Riquelme, H., &Rickards, T. (1992). Hybrid conjoint analysis: an estimation probe in new venture decisions. *Journal of Business Venturing, 7*(6), 505-518.https://doi.org/10.1016/0883-9026(92)90022-J

Schumpeter, J.A. (1934). The theory of economic development: an inquiry into profits, capital, credit, interest, and the business cycle. *Social Science Electronic Publishing, 25*(1), 90-91.

Tan, C.Q., Jiang, Z.Z., &Chen, X.H. (2015). An extended TODIM method for hesitant fuzzy interactive multicriteria decision making based on generalized Choquet integral. *Journal of Intelligent and Fuzzy Systems, 29*(1), 293-305.https://doi.org/10.3233/IFS-151595

Torra, V. (2010). Hesitant fuzzy sets. *International Journal of Intelligent Systems, 25*(6), 529-539.https://doi.org/10.1002/int.20418

Tyebjee, T.T., &Bruno, A.V. (1984). Venture capital: investor and investee perspectives. *Technovation, 2*(3), 185-208.https://doi.org/10.1016/0166-4972(84)90003-8

Wei, C.P., Ren, Z.L., &Rodrǵuez, R.M. (2015). A hesitant fuzzy linguistic TODIM method based on a score function. *International Journal of Computational Intelligence Systems, 8*(4), 701-712.https://doi.org/10.1080/18756891.2015.1046329

Widyanto, H.A., &Dalimunthe, Z. (2015). *Evaluation criteria of venture capital firms investing on Indonesians' SME*. New York: Social Science Electronic Publishing.

Xia, M.M., Xu, Z.S., &Chen, N. (2013). Some hesitant fuzzy aggregation operators with their application in group decision making. *Group Decision and Negotiation, 22*(2), 259-279.https://doi.org/10.1007/s10726-011-9261-7

Xu, Z.S., &Xia, M.M. (2011). Distance and similarity measures for hesitant fuzzy sets. *Information Sciences, 181*(11), 2128-2138.https://doi.org/10.1016/j.ins.2011.01.028

Yazdipour, R. (2011). *Advances in entrepreneurial finance*. New York:Springer.https://doi.org/10.1007/978-1-4419-7527-0

Yu, D.J., Zhang, W.Y., & Xu, Y.J. (2013). Group decision making under hesitant fuzzy environment with application to personnel evaluation. *Knowledge-Based Systems, 52*(6), 1-10. https://doi.org/10.1016/j.knosys.2013.04.010

Yu, D.J., & Li, D.F. (2014). Dual hesitant fuzzy multi-criteria decision making and its application to teaching quality assessment. *Journal of Intelligent and Fuzzy Systems, 27*(4), 1679-1688.https://doi.org/10.3233/IFS-141134

Zacharakis, A.L., Mcmullen, J. S., &Shepherd, D.A. (2007). Venture capitalists' decision making across three countries: an institutional theory perspective. *Journal of International Business Studies, 38*(5), 691-708. https://doi.org/10.1057/palgrave.jibs.8400291

Zhang, X.L., &Xu, Z.S. (2014). The TODIM analysis approach based on novel measured functions under hesitant fuzzy environment. *Knowledge-Based Systems, 61*(2), 48-58.https://doi.org/10.1016/j.knosys.2014.02.006

Zhang, X.L., &Xu, Z.S. (2017). *Hesitant fuzzy methods for multiple criteria decision analysis*(pp. 31-69). Switzerland:Springer.https://doi.org/10.1007/978-3-319-42001-1

Zhu, B., Xu, Z.S., &Xia, M.M. (2012). Dual hesitant fuzzy sets. *Journal of Applied Mathematics,2012*,2607-2645.https://doi.org/10.1155/2012/879629

Zhu, B., &Xu, Z.S.(2017). Probability-hesitant fuzzy sets and the presentation of preference relations.*Technological and Economic Development of Economy*(in press).

Gamelook [online]. 2017. CB Insights: startups failure reasons Top 20 [cited July 2017]. Available from Internet: `http://www.gamelook.com.cn/2014/10/185579`

The Classroom of Tencent [online]. 2017. Tencent website [cited July 2017]. Available from Internet: `https://ke.qq.com`

Netease Open Class [online]. 2017. Netease website [cited July 2017]. Available from Internet: `http://open.163.com/cuvocw`

Yimi Guidance [online]. 2017. Shanghai Yi Mi Education Technology Co., Ltd website [cited July 2017]. Available from Internet: `http://www.1mifudao.com`

Qinxue Education [online]. 2017. Qinxue (Beijing) Network Education Technology Co. Ltd website [cited July 2017]. Available from Internet: `http://www.qinxue100.com`

Puxin Education [online]. 2017. Puxin Education Technology Group Co. Ltd website [cited July 2017]. Available from Internet: `http://www.pxjy.com`